\documentclass{emulateapj}
\usepackage{epsfig}

\begin{document}

\slugcomment{\apj\ Letters, submitted 25 Aug 2014 (Printed \today)}
\title{Cooling Time, Freefall Time, and Precipitation in the Cores of {\em ACCEPT} Galaxy Clusters}
\author{G. Mark Voit\altaffilmark{1} and Megan Donahue\altaffilmark{1}
         } 
\altaffiltext{1}{Department of Physics and Astronomy,
                 Michigan State University,
                 East Lansing, MI 48824, 
                 voit@pa.msu.edu}

\begin{abstract}
Star formation in the universe's largest galaxies---the ones at the centers of galaxy clusters---depends critically on the thermodynamic state of their hot gaseous atmospheres.  Central galaxies with low-entropy, high-density atmospheres frequently contain multiphase star-forming gas, while those with high-entropy, low-density atmospheres never do.  The dividing line between these two populations in central entropy, and therefore central cooling time, is amazingly sharp. Two hypotheses have been proposed to explain the dichotomy.  One points out that thermal conduction can prevent radiative cooling of cluster cores above the dividing line.  The other holds that cores below the dividing line are subject to thermal instability that fuels the central AGN through a cold-feedback mechanism. Here we explore those hypotheses with an analysis of the H$\alpha$ properties of {\em ACCEPT} galaxy clusters.  We find that the two hypotheses are likely to be complementary.  Our results support a picture in which cold clouds inevitably precipitate out of cluster cores in which cooling outcompetes thermal conduction and rain down on the central black hole, causing AGN feedback that stabilizes the cluster core.  In particular, the observed distribution of the cooling-time to freefall-time ratio is nearly identical to that seen in simulations of this cold-feedback process, implying that cold-phase accretion, and not Bondi-like accretion of hot-phase gas, is responsible for the AGN feedback that regulates star formation in large galaxies.
\end{abstract}


\section{Introduction}

\setcounter{footnote}{0}

The hardest thing to get right in cosmological simulations of galaxy clusters is the structure of the intracluster medium (ICM) in cluster cores \cite[e.g.,][]{Nagai+2007ApJ...655...98N, Fabjan+10, Dubois+2011MNRAS.417.1853D, Skory+2013ApJ...763...38S}.   Radiative cooling within the central $\sim 100$~kpc tends to produce a strong cooling flow unless feedback intervenes to replace the radiated energy.  Without strong feedback, rapid star formation makes the simulated central galaxy far too blue and luminous, and the ICM temperature gradient peaks well inside 100~kpc as gas with excessively high entropy flows into the cluster core.  

Observations strongly suggest that outflows from active galactic nuclei (AGNs) provide that feedback \cite[see][and references therein]{mn07, McNamaraNulsen_2012NJPh...14e5023M}.  Clusters with short central cooling times tend to have AGNs that inflate cavities in the ICM at distances $\sim 10-30$~kpc from the center.  The total amount of energy in those cavities is sufficient to replace the radiated energy, at least to an order of magnitude, but the exact mechanism of energy transfer remains unknown.  Clearly the supermassive black hole within the central parsec of a galaxy cluster is responding to the hot-gas environment, and its response is finely tuned to replace the radiated energy without disrupting the cool core itself \cite[e.g.,][]{vd05, Gaspari+2011MNRAS.415.1549G}.  But how does the feedback engine know when to turn on, and how does it determine the correct heating rate?

One of the primary clues to the triggering of AGN feedback in cluster cores emerged from our large {\em Chandra} study of ICM entropy profiles ({\em ACCEPT}), which showed that strong feedback is present only in clusters with low-entropy cores \cite[]{Cavagnolo+08, Cavagnolo+09}.  Almost all of the 239 entropy profiles in {\em ACCEPT} can be fit with the form $K(r) = K_0 \, + \, K_{100} (r/100 \, {\rm kpc})^\alpha$, where $K \equiv kTn_e^{-2/3}$.  Strong feedback is seen primarily in systems with core entropy $K_0 < 30 \, {\rm keV \, cm^2}$, and the exceptions have low-entropy coronae on kiloparsec scales \cite[]{Sun09}.  This core-entropy criterion is roughly equivalent to a central cooling time $\sim 1$~Gyr, in alignment with the finding of \cite{df08} that systems with longer cooling times do not contain X-ray cavities.

Two other phenomena common in clusters with $K_0 < 30 \, {\rm keV \, cm^2}$ are absent from clusters with greater core entropy:  extended emission-line nebulae and star formation in the central cluster galaxy \cite[]{Cavagnolo+08, Rafferty+08, Hoffer+2012ApJS..199...23H}.  The nebulae generally signify the presence of multiphase gas, because systems with H$\alpha$ emission also tend to have abundant molecular gas \cite[]{Edge_2001MNRAS.328..762E}.  Star formation is further confirmation that multiphase gas is present.

These results suggest that strong AGN feedback is triggered by development of a multiphase intracluster medium, implying that accretion of cold clouds \cite[e.g.,][]{ps05} is more important than accretion of hot intercloud gas.  However, most implementations of AGN feedback in cluster simulations assume Bondi-like accretion from the hot phase \cite[e.g.,][]{Sijacki+2007MNRAS.380..877S, Fabjan+10, Dubois+2011MNRAS.417.1853D} .  While Bondi accretion may be energetically plausible in elliptical galaxies \cite[]{Allen+2006MNRAS.372...21A}, it is too feeble to power the larger outflows required in massive clusters \cite[]{Rafferty+2006ApJ...652..216R}.   Furthermore, the cooling time at small radii can be much shorter than the inflow timescale, invalidating the fundamental Bondi assumption of isentropic accretion \cite[]{MathewsGuo_2012ApJ...754..154M}.

The core-entropy threshold for feedback, multiphase gas, and star formation was initially interpreted in terms of conduction.  \cite{Voit+08} showed that inward heat conduction in systems with $K_0 > 30 \, {\rm keV \, cm^2}$ could plausibly replace the energy radiated from the central regions and prevent cooling and condensation \cite[but see][]{Soker_2008ApJ...684L...5S}.  In the conduction interpretation, AGN feedback is not necessary unless $K_0 < 30 \, {\rm keV \, cm^2}$, in which case conduction fails to keep pace with cooling, leading to development of a multiphase medium and cold accretion \cite[]{Voit_2011ApJ...740...28V}.

Alternatively, the multiphase threshold has been interpreted as resulting from thermal instability.  Numerical simulations show that thermal instability can produce a multiphase medium in a spherical potential if the ratio of cooling time ($t_{\rm c}$) to freefall time ($t_{\rm ff}$) is less than $\sim 10$, as long as feedback heating is in approximate global balance with radiative cooling \cite[]{McCourt+2012MNRAS.419.3319M, Sharma+2012MNRAS.420.3174S, Gaspari+2012ApJ...746...94G, LiBryan_2014ApJ...789..153L}.  Heating is necessary for thermal instability because perturbations in a pure cooling flow grow too slowly to produce a multiphase medium if $t_{\rm c} > t_{\rm ff}$ \cite[]{bs89}.   According to the thermal-instability interpretation, globally balanced cluster cores in which $t_{\rm c}/t_{\rm ff} < 10$ should harbor an extended multiphase medium, and this finding was initially supported by analysis of a small cluster sample \cite[]{McCourt+2012MNRAS.419.3319M}.

This paper analyzes a much larger cluster sample to explore how both the core-entropy criterion and the thermal-instability criterion relate to observations of H$\alpha$ emission in cluster cores.  Section 2 updates the results of \cite{Cavagnolo+08} and \cite{Voit+08}, showing that the core-entropy criterion sharply separates clusters with H$\alpha$ emission from those without it. Section 3 expands on the results of \cite{McCourt+2012MNRAS.419.3319M}, showing that the thermal-instability criterion does not produce a clear separation between systems with and without H$\alpha$ emission but instead correlates with the prominence of that emission, and presumably also with the total amount of multiphase gas.  Section 4 discusses the implications of our results for the triggering of AGN feedback. Section 5 considers how gas precipitating out of the ICM can become dusty. Section 6 summarizes the paper.

\section{Core Entropy and H$\alpha$ Emission}
\label{sec-K0_Ha}

\begin{figure}[t]
\includegraphics[width=3.45in, trim = 0.1in 0.2in 0.0in 0.05in]{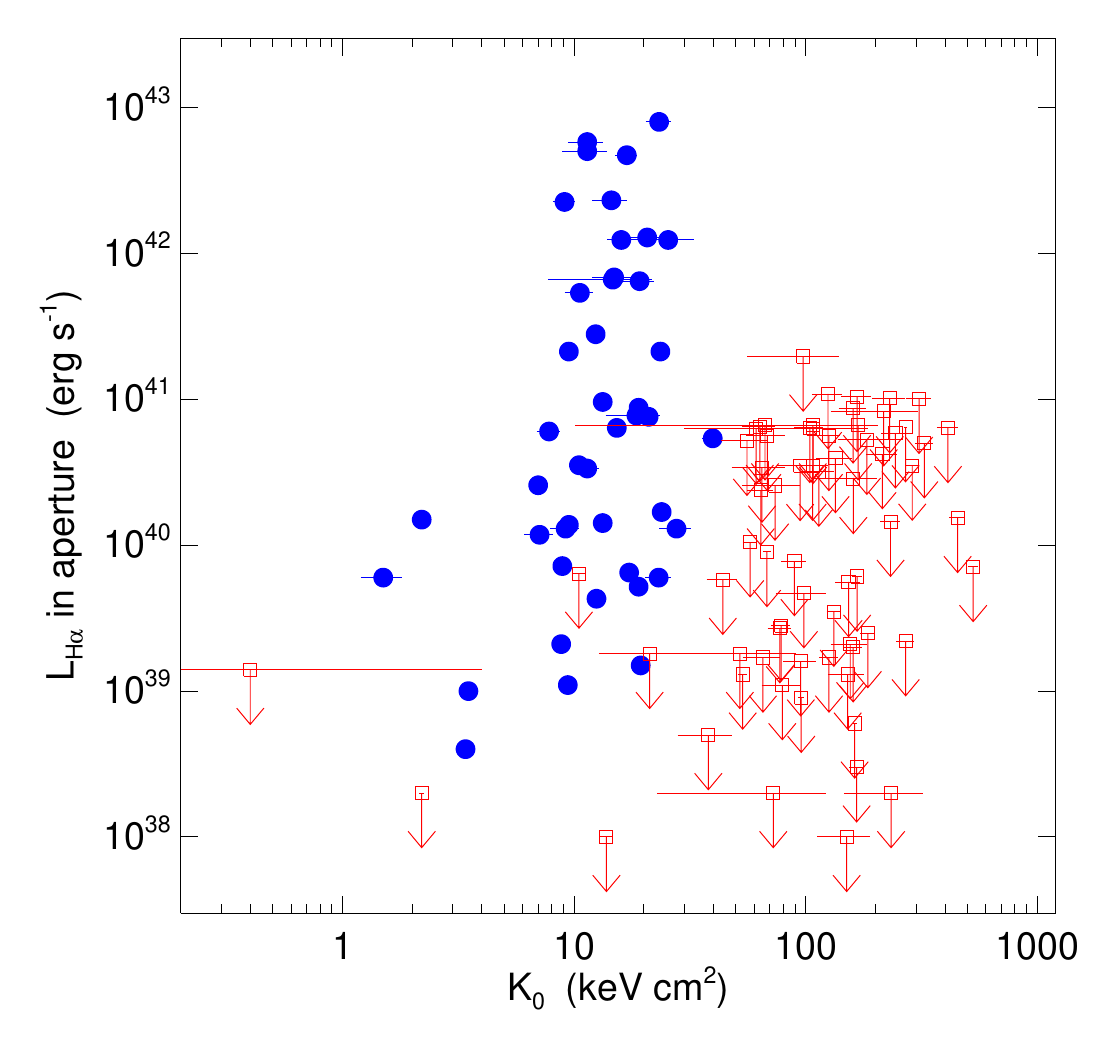} \\
\caption{ \footnotesize 
Relationship between core entropy $K_0$ and H$\alpha$ luminosity, which traces the multiphase medium in cluster cores.  Filled circles (blue) show detections. Empty squares (red) show upper limits.  
\vspace*{1em}
\label{k0-lha}}
\end{figure}

Figure~\ref{k0-lha} shows the relationship between core entropy and core H$\alpha$ luminosity for all {\em ACCEPT} clusters with H$\alpha$ measurements.  Except for a small handful of objects, there is a clear separation in $K_0$ between cluster cores with and without H$\alpha$ emission, consistent with the hypothesis that a multiphase medium develops and triggers AGN feedback when radiative cooling becomes too rapid for conduction to keep pace.  As described in \cite{Cavagnolo+08}, the H$\alpha$ measurements come from a heterogenous set of sources, including many long-slit spectra, meaning that they are often lower limits on the true values.  We have therefore suppressed the error bars on the detections to make the figure more legible.   The main quantitative differences between this figure and Figure~1 of \cite{Cavagnolo+08} are shifts in the horizontal direction having to do with updated $K_0$ values.  Only one cluster with $K_0 > 30 \, {\rm keV \, cm^2}$ has an H$\alpha$ detection (Zwicky 2701), and even that cluster may be consistent with $K_0 \approx 30 \, {\rm keV \, cm^2}$ within the systematic measurement uncertainties \cite[][]{Cavagnolo+09}.

\begin{figure}[t]
\includegraphics[width=3.45in, trim = 0.1in 0.2in 0.0in 0.05in]{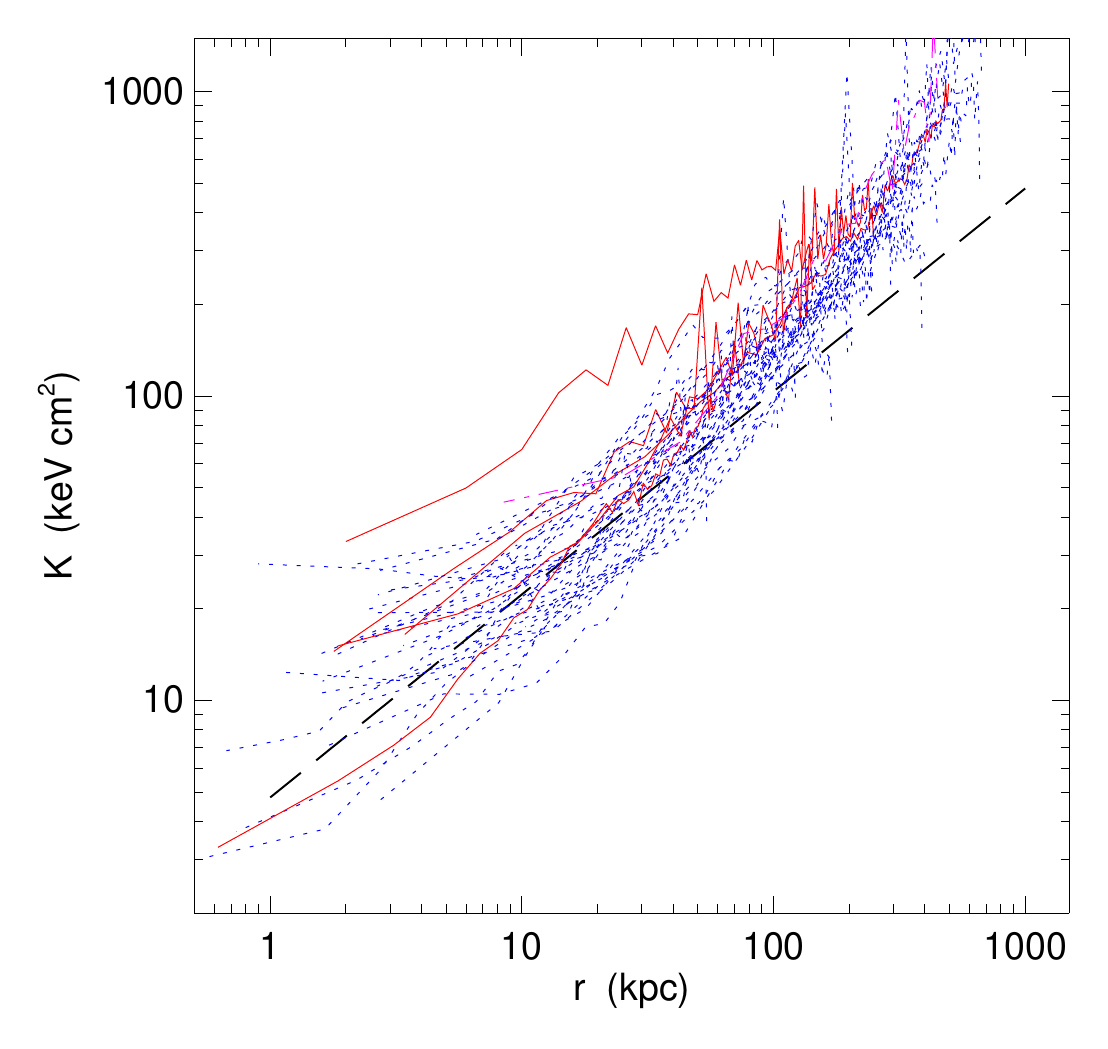} \\
\caption{ \footnotesize 
Entropy profiles for {\em ACCEPT} clusters with $K_0 < 30 \, {\rm keV \, cm^2}$ and H$\alpha$ spectroscopic coverage.  Dotted (blue) lines show clusters with H$\alpha$ emission.  Solid (red) lines show clusters without H$\alpha$ detections.   Electron thermal conduction (with $f_c = 1/3$) can offset radiative cooling in clusters with profiles above the dashed (black) line.  
\vspace*{0em}
\label{kprofs-ha}}
\end{figure}

Five objects with $K_0 < 30 \, {\rm keV \, cm^2}$ have no detectable H$\alpha$ emission:  Abell~2029, Abell~2107, Abell~2151, EXO~0422-086, and RBS~0533.  Figure~\ref{kprofs-ha} shows that conduction can potentially prevent gas from condensing in the cores of four of them.  The figure illustrates the $K(r)$ entropy profiles for all of the objects in Figure~\ref{k0-lha} with $K_0 < 30 \, {\rm keV \, cm^2}$.  Dotted (blue) lines represent profiles of systems with core H$\alpha$ emission, and solid (red) lines represent profiles of systems without H$\alpha$.  The (black) dashed line shows the critical $K(r)$ profile for conductive balance given by \cite{Voit_2011ApJ...740...28V} for a conduction suppression factor $f_c = 1/3$ applied to the full \cite{Spitzer62} conductivity.  In clusters with profiles above this line, thermal conduction in the radial direction can supply enough heat to the core to compensate for radiative cooling.  The figure shows that four of the five clusters without H$\alpha$ emission have $K(r)$ profiles that remain above the critical line.  Only one, RBS~0533, has an entropy profile that drops below it.

\section{Cooling Time, Freefall Time, and H$\alpha$ Emission}
\label{sec-K0_tff_tc}

\begin{figure}[t]
\includegraphics[width=3.45in, trim = 0.1in 0.2in 0.0in 0.05in]{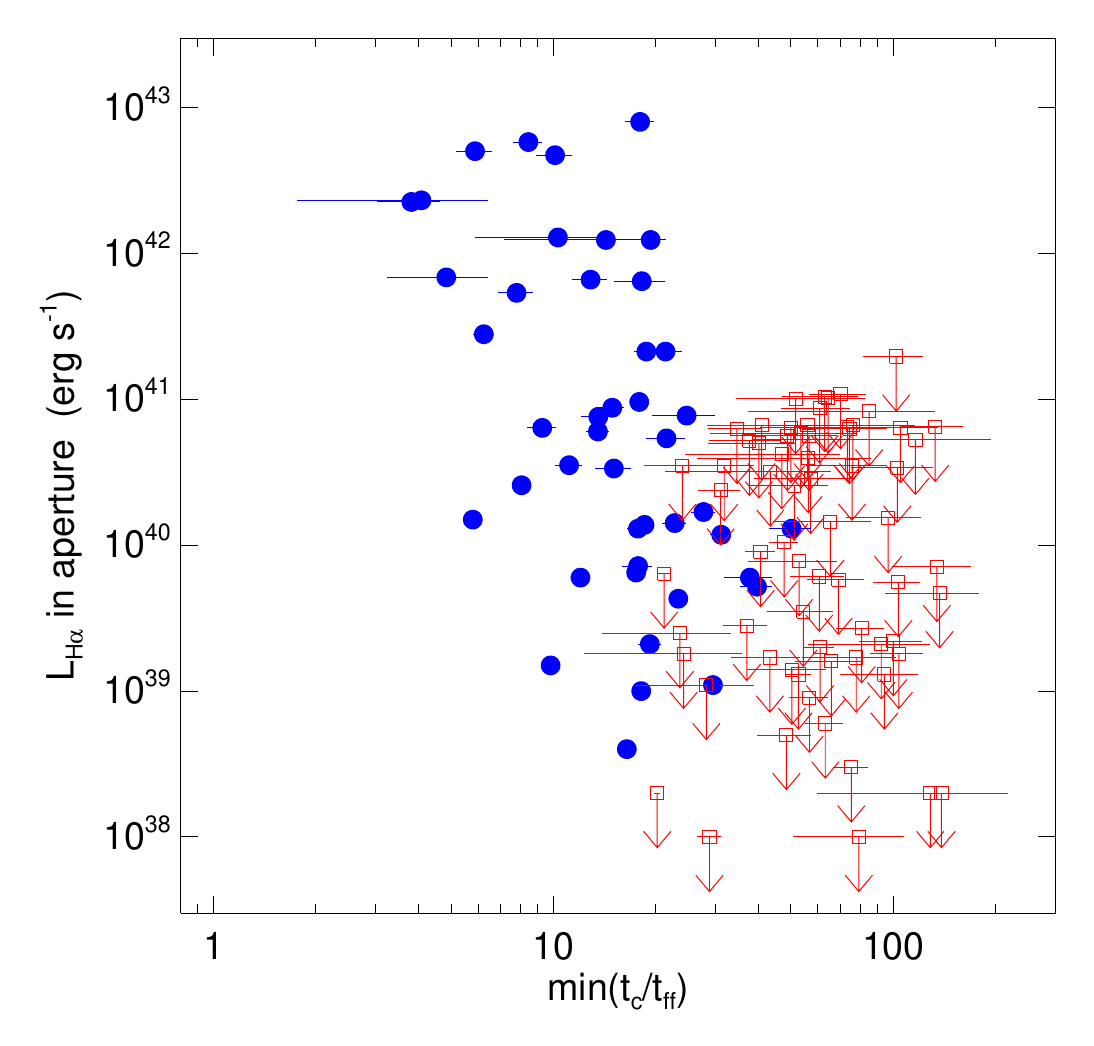} \\
\caption{ \footnotesize 
Relationship between the minimum value of $t_{\rm c}/t_{\rm ff}$ and H$\alpha$ luminosity. Blue circles show H$\alpha$ detections and red squares show upper limits.  
\vspace*{0em}
\label{tc-lha}}
\end{figure}

The relationship between H$\alpha$ emission and the minimum value of $t_{\rm c}/t_{\rm ff}$ in cluster cores is somewhat different than the one for core entropy.  Figure~\ref{tc-lha} shows that relationship. There is not a clear threshold at $t_{\rm c}/t_{\rm ff} \approx 10$ like the one at $K_0 \approx 30 \, {\rm keV \, cm^2}$, indicating that the presence of a multiphase medium depends on more than just thermal instability.  However, the steep anticorrelation of $L_{{\rm H}\alpha}$ with min($t_{\rm c}/t_{\rm ff}$) does suggest that thermal instability determines the {\em amount} of multiphase gas in the core and therefore the strength of the feedback reponse owing to accretion of cold, condensed gas into the AGN.

In order to make Figure~\ref{tc-lha}, we needed to estimate the gravitational potential as a function of radius for each cluster, which required a few assumptions.  First, we assumed hydrostatic equilibrium, allowing us to estimate the gravitational potential from the gas pressure and temperature profiles in the {\em ACCEPT} catalog.  Second, we smoothed the pressure and temperature profiles by fitting $\ln P(r)$ and $\ln T(r)$ with third-order polynomials in $\ln r$.  Third, we assumed that a large galaxy resides at the center of each halo.  This last assumption is important for three reasons: (1) nearly all cool cores are centered on a cluster's brightest galaxy, (2) that galaxy's stars dominate the gravitational potential within the central $\sim 10$--20~kpc, and (3) flattening of the pressure profile in a cluster core can lead to a poorly determined or even unconstrained X-ray measurement of the potential at small radii. Given the paucity of data on the velocity dispersions of these central galaxies, we approximated them all as singular isothermal spheres with a 250 km~s$^{-1}$ velocity dispersion \cite[consistent with e.g.,][]{Bernardi+2007AJ....133.1741B}.  For clusters without large galaxies at their centers, this last assumption is incorrect.  However, such clusters do not have cool cores and therefore have values of $t_{\rm c}/t_{\rm ff}$ so large that any inaccuracy introduced by this faulty assumption is inconsequential for our analysis.

Our assumption of hydrostatic equilibrium is likely to be inaccurate, but probably not by more than a few tens of percent \cite[e.g.,][]{Churazov+2010MNRAS.404.1165C}.  Inaccuracies arising from a lack of hydrostatic equilibrium will lead to underestimates of the gravitational potential and overestimates of $t_{\rm ff}$, meaning that some of our $t_{\rm c}/t_{\rm ff}$ values may be a few tens of percent too low.  However, assuming that a large galaxy dominates the gravitational potential at the very center of a cluster mitigates these uncertainties.

\section{Multiphase Gas and AGN Triggering}
\label{sec-triggering}

This section interprets what our results imply about cooling and condensation of the intracluster medium and triggering of the AGN feedback presumed to limit it.  First we consider the significance of the core-entropy dichotomy.  Then we turn to the significance of the dependence of H$\alpha$ emission on the minimum value of $t_{\rm c}/t_{\rm ff}$.

Low core entropy looks like a necessary but not sufficient condition for development of a multiphase medium in a cluster core.  We therefore surmise that high core entropy {\em prohibits} the existence of a multiphase medium.  Elsewhere we have proposed that electron thermal conduction with a Spitzer suppression factor $\sim 1/3$ is what prohibits a multiphase medium in this case \cite[]{Donahue+05, Voit+08, Voit_2011ApJ...740...28V}.  Theoretical models support this viewpoint \cite[]{gor08}, as long as the level of turbulence in the core is sufficient to keep the suppression factor from dropping too far below $\sim 1/3$ \cite[]{pqs10}.  However, others have argued that MHD effects produce far greater suppression \cite[e.g.,][]{Schekochihin+2008PhRvL.100h1301S}, and there is some observational support for strong suppression \cite[]{GaspariChurazov_2013A&A...559A..78G}.  For that reason, turbulent heat diffusion (which can be considered a generalized heat conduction mechanism) also needs to be investigated as potentially responsible for the $K_0$ dichotomy.

We would like to stress that the mechanism responsible for suppressing multiphase structure in cores with $K_0 > 30 \, {\rm keV \, cm^2}$ must do more than just stop condensation of the intracluster medium.  It must also transport the gas injected by normal stellar mass loss from the central galaxy's stars away from the cluster core while heating it to the ambient ICM temperature and destroying its dust grains \cite[]{VoitDonahue_2011ApJ...738L..24V}.  And it must accomplish this task without producing obvious evidence for AGN feedback, since cluster cores with $K_0 > 30 \, {\rm keV \, cm^2}$ seem not to harbor strongly radio-emitting AGN \cite[]{Cavagnolo+08} unless those AGNs reside in cool kpc-scale coronae with $K_0 \ll 30 \, {\rm keV \, cm^2}$ \cite[]{Sun09}.  

The amount of heat transport required to heat and remove the ejected stellar gas from the central galaxy in a cluster with $K_0 > 30 \, {\rm keV \, cm^2}$ depends somewhat on the temperature of the ambient hot gas but is typically at least as large as the heat input needed to keep the core gas from cooling.  For example, consider a 5 keV cluster core with $K_0 \sim 30 \, {\rm keV \, cm^2}$.  The mass of ambient gas within 30~kpc of the center is $\sim 3 \times 10^9 M_\odot$, and the central galaxy's stars are shedding $\sim 3 M_\odot \, {\rm yr}^{-1}$ in that same region, requiring a heating+transport timescale $\sim 1$~Gyr, essentially the same as the cooling time of the ambient gas.  In clusters of greater central entropy, the energy input required to heat and transport the ejected stellar gas {\em exceeds} the cooling luminosity of the ambient gas.

This transport mechanism apparently falters in cluster cores below $\sim 30$~keV~cm$^2$, allowing a multiphase medium to persist, but not in all cases.  Recent theoretical work has clarified the conditions under which condensation of a low-entropy core can produce multiphase structure and indicates that the threshold for precipitation in a spherical potential is $t_{\rm c}/t_{\rm ff} \approx 10$ \cite[]{Sharma+2012MNRAS.420.3174S, Gaspari+2012ApJ...746...94G}.  However, according to Figure~\ref{tc-lha}, this condition is sufficient but not necessary.  Cores with H$\alpha$ emission span the range $\min(t_{\rm c}/t_{\rm ff}) \approx 5$--20, with a few ranging up to ratios exceeding 30.

How are we to interpret the multiphase structure in cluster cores with $\min(t_{\rm c}/t_{\rm ff}) > 10$?  Recent simulations of AGN feedback triggered by cold precipitation provide some important clues \cite[see][]{Gaspari+2012ApJ...746...94G, Gaspari+2013MNRAS.432.3401G, Gaspari+2014arXiv1407.7531G, LiBryan_2014ApJ...789...54L, LiBryan_2014ApJ...789..153L}.  In cores without multiphase gas, the ambient medium needs to drop below $t_{\rm c}/t_{\rm ff} \approx 10$, and maybe as low as $t_{\rm c}/t_{\rm ff} \approx 3$ \cite[]{LiBryan_2014ApJ...789..153L}, before a multiphase medium develops and fuels the AGN.  When the AGN responds, it heats and inflates the ambient medium, raising its entropy and lowering its density until $t_{\rm c}/t_{\rm ff} \gtrsim 10$, but does not eradicate the multiphase medium.  Instead, fluctuations in the cold accretion rate lead to fluctuations in AGN feedback, causing $\min (t_{\rm c}/t_{\rm ff})$ to vary between $\sim 4$ and $\sim 20$ \cite[see Figure 10 of][]{Gaspari+2012ApJ...746...94G}, the same range occupied by most of the multiphase cluster cores in Figure~\ref{tc-lha}.  This agreement should be considered strong observational support for the precipitation-driven cold feedback mechanism.

\begin{figure}[t]
\includegraphics[width=3.45in, trim = 0.1in 0.2in 0.0in 0.05in]{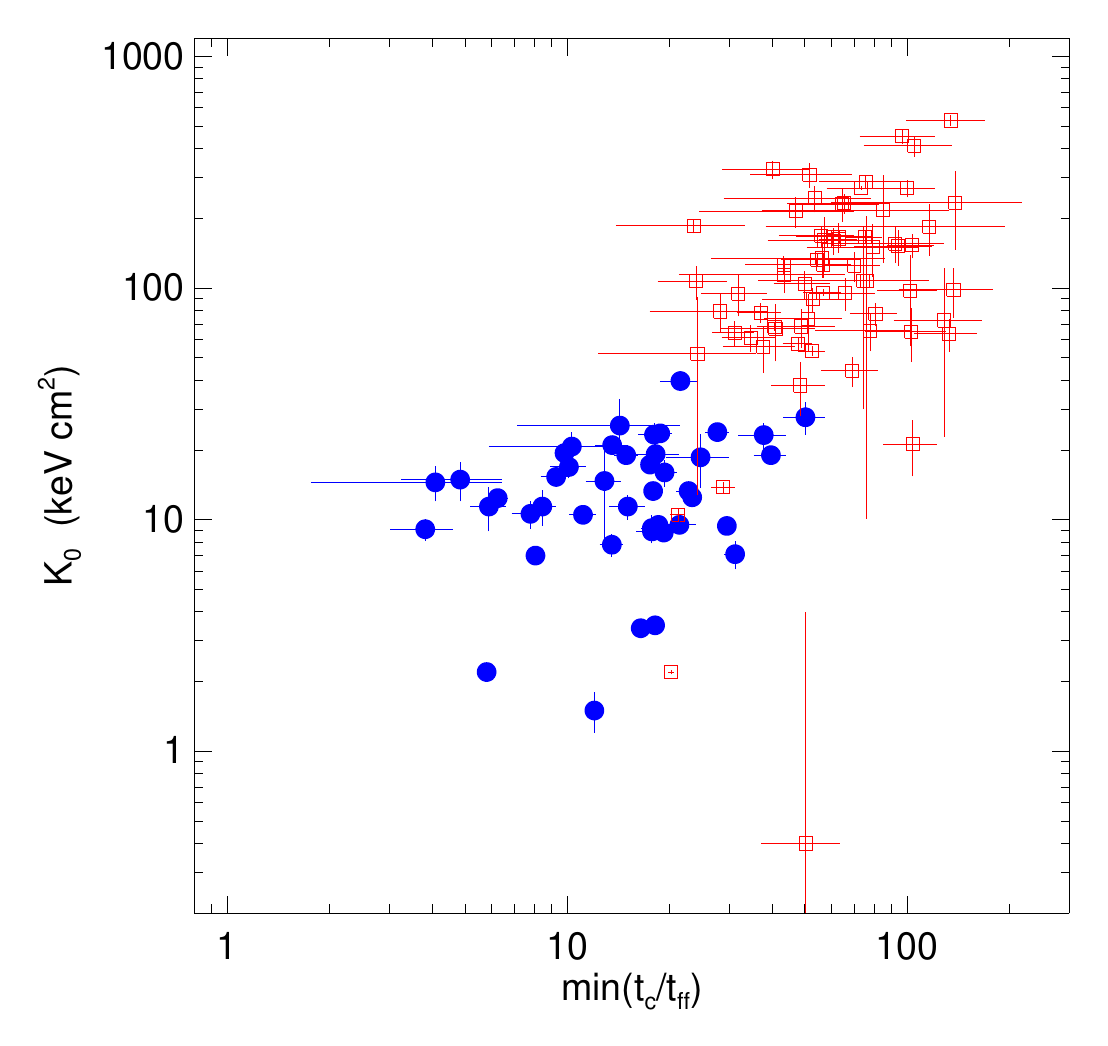} \\
\caption{ \footnotesize 
Relationship between the minimum value of $t_{\rm c}/t_{\rm ff}$ and and core entropy $K_0$. Blue circles show systems in which H$\alpha$ is detected and red squares show systems in which it is not.  \vspace*{0em}
\label{mintrat-k0}}
\end{figure}

Figure~\ref{mintrat-k0} provides some insight into the exceptions.  All of the H$\alpha$-emitting cores with $\min (t_{\rm c}/t_{\rm ff}) > 20$ remain below $K_0 \approx 30 \, {\rm keV \, cm^2}$ indicating that the cold-gas eradication mechanism operating in clusters of higher core entropy is not quite strong enough to eliminate their multiphase structure.  We suspect these cores are examples of cases in which a strong AGN outburst originally triggered during a low $\min(t_{\rm c}/t_{\rm ff})$ state has temporarily boosted the core entropy, but not high enough for heat transport from outside the core to eliminate the multiphase gas.  In this interpretation, the presence of H$\alpha$ in these cores is a hysteresis effect in the AGN feedback cycle.

Figure~\ref{mintrat-k0} also suggests a natural explanation for the five low-$K_0$ clusters without evidence for multiphase gas.  While it remains possible that four of them are stabilized by conduction (see Figure~\ref{kprofs-ha}), it is also the case that all five have $\min(t_{\rm c}/t_{\rm ff}) > 20$, suggesting that this timescale ratio has not yet dropped low enough to produce precipitation in these cores. 

\section{Precipitation, Dust, Advection, \& Mixing}

The preceding sections present a strong case for precipitation-driven AGN feedback as the mechanism that stabilizes the lowest-entropy cluster cores, but a fly remains in the ointment: The multiphase gas in cluster cores is known to be dusty \cite[e.g.,][]{Sparks+1989ApJ...345..153S, DonahueVoit_1993ApJ...414L..17D, Donahue+2011ApJ...732...40D, Rawle+2012ApJ...747...29R}, yet the surrounding hot medium is virtually dust-free \cite[e.g.,][]{Bovy+2008ApJ...688..198B}, especially near the center \cite[]{McGeeBalogh_2010MNRAS.405.2069M}. In fact, sputtering at the center of a cool-core cluster should destroy all the dust grains in less than a million years \cite[based on ][]{DraineSalpeter_1979ApJ...231...77D}.  So how can the precipitated gas be dusty?

One potential solution to this dilemma is feedback-induced mixing.  In the cold regions of the Milky Way's interstellar medium, pre-existing dust grains can grow by accreting refractory elements from the gas phase \cite[]{Draine_2009ASPC..414..453D}.  The most plausible sources of pre-existing dust grains in cluster cores are dying stars in the central galaxy \cite[e.g.,][]{VoitDonahue_2011ApJ...738L..24V}.  Mixing of this stellar dust into cold gas that is accumulating in the central galaxy can preserve it against sputtering, providing sites for molecule formation and further depletion of refractory elements out of the gas phase.  Such a mechanism for incorporation of gas-phase metals from the ICM into dust grains is far easier to understand than {\em ab initio} dust formation in dust-free clouds of condensing ICM gas, particularly since those clouds are unlikely even to form molecules until they have first formed dust grains. Dust in extended multiphase ICM filaments therefore indicates that those filaments consist of gas that has been pulled outward from smaller radii, either in the sheaths around AGN jets or in the wakes of buoyant bubbles.  

If this is the correct interpretation of dust in cluster cores, then it implies that outward advection of low-entropy gas promotes precipitation, perhaps explaining why $t_{\rm c}/t_{\rm ff}$ values in multiphase cluster cores are sometimes  greater than those seen in idealized simulations.  As jets and bubbles lift adjacent low-entropy gas to higher altitudes of greater ambient entropy, they locally decrease the $t_{\rm c}/t_{\rm ff}$ ratio in the uplifted lower-entropy gas.  Radial mixing of stratified gas that is marginally stable to precipitation therefore destabilizes it \cite[see e.g.,][]{Revaz+2008A&A...477L..33R, LiBryan_2014ApJ...789..153L}.

Mixing of dust into the precipitating gas can further destablize it, because the cooling rate of a hot, dusty medium is far greater than that of a hot, dust-free medium at the same temperature \cite[e.g.,][]{DwekArendt_1992ARA&A..30...11D}.  Sputtering rapidly destroys some of the dust, but the temporary decrease in cooling time might be sufficient to cause precipitation in cases where it otherwise wouldn't happen.  If mixing of dust into the ambient gas does indeed catalyze precipitation, then one would expect precipitating gas to be dustier, on average, than gas that is not precipitating.  

\section{Summary}
\label{sec-summary}

We have used the subset of {\em ACCEPT} galaxy clusters with published H$\alpha$ observations to evaluate two hypotheses for the origin of multiphase gas in cluster cores, one based on conduction and $K_0$ and another based on thermal instability and $t_{\rm c}/t_{\rm ff}$ .  Our results suggest that these two hypotheses are complementary.  The dichotomy between homogeneous cores and multiphase cores, as traced by H$\alpha$ emission, is most distinct when plotted as a function of the core-entropy parameter $K_0$.  None of the {\em ACCEPT} clusters with high core entropy has measurable H$\alpha$ emission, indicating that low core entropy is a {\em necessary} condition for a multiphase core.  On the other hand, all of the {\em ACCEPT} clusters with $\min(t_{\rm c}/t_{\rm ff}) < 20$ have H$\alpha$ emission, indicating that this is a {\em sufficient} condition for a multiphase core.

Exceptions to the inverse relations support the notion of complementarity. The few clusters with low core entropy and no H$\alpha$ emission all have $t_{\rm c}/t_{\rm ff} > 20$, suggesting that their central cooling times are not short enough for precipitation (although four of the five could also be stabilized by thermal conduction).  And all of the H$\alpha$ emitting clusters with $t_{\rm c}/t_{\rm ff} > 20$ have low core entropy, consistent with the idea that heat transport is incapable of eliminating the multiphase medium in a low-entropy core, once it has formed.  We suspect that these clusters are examples of hysteresis in the AGN feedback cycle, which can intermittently raise the central cooling time so that occasional excursions to $t_{\rm c}/t_{\rm ff} > 20$ occur once a multiphase medium has developed.

Most multiphase cores have $4 < \min (t_{\rm c}/t_{\rm ff}) < 20$. This is the same interval spanned over time in cluster-core simulations that rely on precipitating clouds to fuel AGN feedback.  We consider this agreement strong evidence in favor of the cold-precipitation paradigm for stabilizing the lowest entropy cluster cores. 

The most serious problem with the precipitation paradigm is the dustiness of the multiphase gas.  We suggest that the presence of dust indicates that the precipitating gas has been drawn out of the central galaxy by jets or buoyant bubbles after it has mixed with dusty ejecta from the central galaxy's dying stars.  Both uplift and the introduction of dust promote precipitation, because both reduce $t_{\rm c}/t_{\rm ff}$ in the uplifted gas.

\vspace*{1.0em}

The authors thank M. Gaspari, M. McCourt, P. Sharma for helpful conversations and NSF for support through grant AST-0908819.  


\end{document}